\documentclass[prx,twocolumn,superscriptaddress,nolongbibliography]{revtex4-2}
\usepackage{graphicx}
\usepackage{bm}
\usepackage{amssymb}
\usepackage{color}
\usepackage{amsmath}
\usepackage{mathrsfs} 
\usepackage{amstext}
\usepackage[polish,english]{babel}
\usepackage{latexsym}
\usepackage{array}             
\usepackage{multirow}         
\usepackage[usenames,dvipsnames]{xcolor}
\usepackage[colorlinks=true,citecolor=Blue,linkcolor=RubineRed,urlcolor=Blue]{hyperref}
\usepackage{tocloft}

\usepackage{cancel}
\usepackage[normalem]{ulem}

\usepackage{float}

\def\la{\lambda}

\def\N{{\rm N}}

\def\la{\langle}
\def\ra{\rangle}
\def\om{\omega}

\newcommand{\beq}{\begin{equation}}
\newcommand{\eeq}{\end{equation}}
\newcommand{\beqa}{\begin{eqnarray}}
\newcommand{\eeqa}{\end{eqnarray}}

\begin{document}

\title{Exact Delta Kick Cooling,  Time-Optimal Control  of Scale-Invariant Dynamics, and  Shortcuts to Adiabaticity Assisted by Kicks}
\author{L\'eonce Dupays}
\affiliation{Department  of  Physics  and  Materials  Science,  University  of  Luxembourg,  L-1511  Luxembourg, G. D.  Luxembourg}
\affiliation{Donostia International Physics Center,  E-20018 San Sebasti\'an, Spain}
\author{David C. Spierings}
\affiliation{Centre for Quantum Information and Quantum Control and Institute for Optical Sciences, Department of Physics, University of Toronto, Toronto, Ontario, Canada}
\author{Aephraim  M.  Steinberg}
\affiliation{Centre for Quantum Information and Quantum Control and Institute for Optical Sciences, Department of Physics, University of Toronto, Toronto, Ontario, Canada}
\author{Adolfo del Campo}
\affiliation{Department  of  Physics  and  Materials  Science,  University  of  Luxembourg,  L-1511  Luxembourg, G. D.  Luxembourg}
\affiliation{Donostia International Physics Center,  E-20018 San Sebasti\'an, Spain}

\begin{abstract}
Delta kick cooling (DKC) is used to  compress the momentum distribution of ultracold quantum matter. It combines expansion dynamics with the use of kick pulses, designed via classical methods, that bring the system to rest. We introduce an exact approach to DKC  for  arbitrary scale-invariant dynamics of  quantum gases, lifting  the original restrictions to free evolution and noninteracting systems, to account for the control of atomic clouds in a time-dependent harmonic trap that can be either  repulsive (inverted) or confining. We show that DKC assisted by a repulsive potential outperforms the conventional scheme, and that sudden trap-frequency quenches combined with DKC are equivalent to time-optimal bang-bang protocols. We further show that reverse engineering of the scale-invariant dynamics under smooth trap-frequency modulations can be combined with DKC to introduce a new class of shortcuts to adiabaticity assisted by kicks.
\end{abstract}

\maketitle

Delta kick cooling (DKC)  is a technique for preparing states with low kinetic temperatures (narrow momentum distributions). It relies on unitary dynamics generated by a time-dependent Hamiltonian and it thus preserves the von Neumann entropy of the  system. As a result, it belongs to the family of techniques that achieve ``cooling'' preserving phase-space density \cite{Chu86,Ammann97,Morinaga99,Myrskog00}.
DKC can be broadly used for the cooling of cold atoms \cite{Morinaga99,Marechal99,Myrskog00,Aoki06,Kovachy15,Luan18}, molecules, and ions. It has found applications in interferometry and gravimetry with Bose-Einstein condensates \cite{Muntinga13,Heine2020}. While originally conceived  in a time-of-flight setting \cite{Chu86}, variants of it involve pulsing optical lattices and other potentials \cite{Dong04}.

An alternative approach to  cooling preserving phase-space density relies on 
 expansions engineered  in isolated systems confined in a time-dependent trap \cite{Salamon09,Chen10,Muga09,Stefanatos10,Hoffmann11,delcampo11,Choi2011,Stefanatos11,Choi2011b,delcampo12,Choi12,Stefanatos12,Jarzynski13,delcampo13,Choi13,Deffner14,Patra17,Patra21,delcampo2021probing}. The modulation of the trapping frequencies in such expansions can involve  sudden, abrupt changes 
 as in time-optimal bang-bang protocols. Alternatively, smooth modulations can be  designed using  shortcuts to adiabaticity (STA) \cite{Torrontegui13,delCampo2019} (to be distinguished from recent generalizations to open quantum systems that alter the phase-space density \cite{Dann19,Villazon19,Dupays20,Alipour20,DupaysChenu20}). 
 Such STA have been applied to the cooling of ultracold atoms, including thermal clouds \cite{Schaff10}, Bose-Einstein condensates \cite{Schaff11,Schaff11njp,Zhou2018}, low-dimensional gases \cite{Rohringer2015}, and Fermi gases in the non-interacting and strongly-interacting regimes \cite{Deng18,Deng18Sci,Diao18,delCampo2018}.
 Though most works consider harmonic confinement, it is worth emphasizing that  these techniques are  also applicable to  anharmonic trapping geometries  \cite{delcampo11,delcampo12,Choi13,Jarzynski13,delcampo13,Choi13,Deffner14}. The latter can be explored in ultracold gases -- e.g. using time-averaged potentials \cite{Henderson09} or  digital micromirror devices \cite{Dudley03} -- as well as trapped ions, among other platforms.

In this work, we introduce an exact framework to describe DKC using scale-invariant dynamics of quantum gases, from which the conventional classical analysis is derived as an approximation. 
As a result, we generalize DKC in the presence of interactions and an arbitrary time-dependent harmonic trap, identifying the exact pulse parameters to bring the trapped atomic cloud to rest.
We prove the equivalence between the generalized DKC and bang-bang protocols in time-optimal  control with unbounded frequencies, find exact DKC protocols with realistic  pulses of finite duration and strength, and identify the conditions for the instantaneous-pulse description to hold.
We also introduce a new class of shortcuts to adiabaticity that exploit a combination of reverse-engineering the scale-invariant dynamics and $\delta$-kicks.

\section{DKC and scale invariance}
\begin{figure}[t]
\includegraphics[width=0.95\linewidth]{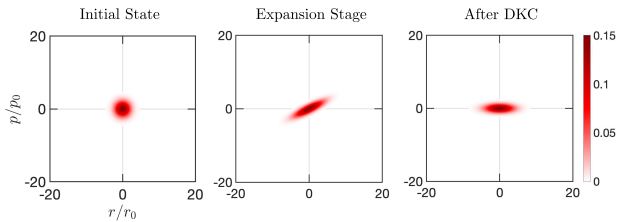}
\caption{Wigner function of a thermal state of a quantum oscillator during a standard DKC protocol. 
During expansion in free space the initial momentum width is conserved as the width in position grows, and in the long-time limit momentum becomes linearly correlated with position.  
A harmonic kick rotates the distribution onto the position axis, lowering the temperature of the cloud.   The initial inverse temperature $\beta_0=1/(\hbar\om_0)$ and the expansion time $t_k=3/2\omega_0$, where $\omega_0$ is the initial trap frequency. Further, $p_{0}=\sqrt{(\hbar m\omega_{0})/2}$ and $r_{0}=\sqrt{\hbar/(2 m \omega_{0})}$.
\label{phase_space}}
\end{figure}

DKC of an atomic cloud admits an intuitive description in the absence of interparticle interactions using classical equations of motion \cite{Chu86,Ammann97,Morinaga99,Myrskog00}. The width  of a wavepacket with initial size $\Delta r_0$ and momentum dispersion $\Delta p_0$ grows in time according to $\Delta r(t) \approx \Delta p_0 t/m=b(t)\Delta r_0$ in the absence of any confining potential.
The dynamics are governed by the kinetic energy and are thus Hamiltonian. Liouville's theorem guarantees that the phase space density is conserved. To bring the expanding cloud to rest, DKC makes use of a pulsed harmonic potential $V=\frac{1}{2}m\om_k^2r^2$ for a short duration $\tau_k$. This pulse acts as a harmonic lens, applying a force $\vec{f}=-m\om_k^2\vec{r}$. The resulting kick alters the momentum of the wavepacket according to $\delta \vec{p}=\vec{f}\tau_k=-m\tau_k\om_k^2\vec{r}$. Particles with momentum $m\vec{r}/t_k$ are brought to rest by choosing $\tau_k\om_k^2=1/t_k$, where $t_k$ is the instant at which the free expansion is interrupted by the kick. In doing so, DKC narrows the momentum distribution by a factor of $\Delta r_0/\Delta r(t)=1/b(t)$, as illustrated in Fig. \ref{phase_space}.

In what follows we adopt a general framework for quantum systems undergoing self-similar dynamics, also known as scale-invariant evolution.
This kind of dynamics is familiar  in the description of  wavepackets in a time-dependent harmonic trap \cite{LewisRiesenfeld69,Lohe08}. As such, it describes as well noninteracting Bose and Fermi gases.
Yet, scale invariance also governs the dynamics of interacting systems, such as one-dimensional gases in the Tonks-Girardeau regime \cite{MinguzziGangardt05,delcampo08},  two-dimensional Bose-Einstein condensates with contact interactions \cite{PR97}, and the three dimensional unitary Fermi gas \cite{Menotti02,Castin12}, among other examples \cite{Gritsev10,delcampo11}. In the Thomas-Fermi regime, it actually governs the evolution of Bose-Einstein condensates in any spatial dimension \cite{Muga09}.
Scale invariance is at the core of time-of-flight imaging techniques, providing the means to reconstruct in-situ properties of a confined atomic cloud, such as the momentum distribution,  from  images of its spatial density after  its release \cite{CastinDum96,Kagan96,Dalfovo99,Giorgini08}. Conversely, scale invariance can also be used to probe properties during time evolution, such as the mean energy and its fluctuations,  when equilibrium properties are known \cite{Jaramillo16,Beau20,delcampo2021probing}.  As we shall see, scale invariance can be used to describe DKC exactly, without approximations, and to generalize it in the presence of time-dependent confinements.

Consider the family of  time-dependent Hamiltonians
\begin{eqnarray}
 H(t)=\sum_{i=1}^{\N}\left[\frac{\vec{p}_i\,^2}{2m}+\frac{1}{2}m\omega(t)^2 \vec{r}_i\,^{2}\right]+\sum_{i<j}V(\vec{r}_i-\vec{r}_j),
 \label{hscale}
\end{eqnarray}
describing $\N$ particles in an isotropic harmonic trap. Particles interact with each other through a homogeneous pairwise potential  fulfilling $V(\lambda \vec{r})=\lambda^{-2}V(\vec{r})$.  Thanks to this scaling property, the dynamics are self-similar, i.e., scale invariant \cite{Castin04,Gritsev10,delcampo11}, a familiar feature in  Bose-Einstein condensates \cite{Kagan96,CastinDum96}.
An energy eigenstate $\Psi(0)$ of the Hamiltonian at $t=0$  with eigenvalue $E(0)$ evolves into
\beqa
\label{psit}
\Psi\left(t\right)&=&
\frac{1}{b^{\frac{D\N}{2}}}\exp\left[i\frac{m\dot{b}}{2\hbar b}\sum_{i=1}^\N \vec{r}_i\,^2-i\int_{0}^t\frac{E(0)}{\hbar b(t')^2}dt'\right]\nonumber\\
& & \times \Psi\left(\frac{\vec{r}_1}{b},\dots,\frac{\vec{r}_\N}{b},t=0\!\right)\,,
\eeqa
where  $D$ denotes the spatial dimension and $b(t)>0$ is the scaling factor that determines the variation of the atomic cloud size. The specific  time-dependence of the latter following an arbitrary modulation of the trapping frequency 
$\om(t)$ can be found by solving the Ermakov equation, 
\beqa 
\label{eq:ermakov}
\ddot{b}+\om(t)^2b=\om_0^2/b^{3},
\eeqa 
with the  boundary conditions $b(0)=1$ and $\dot{b}(0)=0$, as $\Psi(0)$ is assumed to be stationary for $t<0$.
For instance, in time-of-flight (TOF) imaging, the typical expansion factor after suddenly switching off the trap (so that $\om(t)=0$ for $t\geq 0$) is characterized by
\beqa
b_{\rm TOF}(t)=\sqrt{1+\om_0^2t^2}.
\label{btof}
\eeqa
By contrast, in the adiabatic limit, setting $\ddot{b}\approx 0$ in the Ermakov equation, one finds
\beqa
b_{\rm ad}=\sqrt{\frac{\om_0}{\om(t)}}.
\eeqa

We next leverage the description of DKC to scale-invariant dynamics, thus allowing for an exact treatment with no approximations.
To this end, consider an initial equilibrium state of $H(t)$ at $t=0$ when the trap  frequency is $\om(0)=\om_0$.  A  modulation in time  of the trap frequency $\om(t)$ induces a nonequilibrium dynamics of the state of the system described by Eq. (\ref{psit}).
In principle, one can match at a given instant $t_F$ the width of the atomic cloud out of equilibrium with that of a harmonic trap  of  frequency 
\beqa
\label{omF}
\om_F=\frac{\om_0}{b_F^2},
\eeqa
whose ground state has the same width.
With respect to the associated Hamiltonian $H(t_F)$  with $\om(t_F)=\om_F$,
excitations under self-similar dynamics are encoded in the phase factor proportional to $\dot{b}/b$ in Eq. (\ref{psit}). It can be shown that this phase modulation is responsible for the broadening of the momentum distribution, see Appendix \ref{AppendixHeis}.
The cancellation of this phase factor  can be achieved by an instantaneous $\delta$-kick, described by the ``kicked'' Hamiltonian
\beqa
\label{eq:delta}
H_k(t)=H(t)+\delta(t-t_k)\frac{1}{2}m\om_k^2\sum_{i=1}^{\N} \vec{r}_i\,^{2}.
\eeqa
It is well known that the propagator associated with such a Hamiltonian can be decomposed
as
\beqa
U_{ \delta}(t,0)=U(t,t_k+\tau_k)e^{-i\tau_k\frac{m\om_k^2}{2\hbar}\sum_{i=1}^{\N} \vec{r}_i\,^{2}}U(t_k,0),
\eeqa
where $U(t,t')$ is the propagator associated with $H(t)$, and $\tau_k$ is a small time scale during which the kick is applied (finite pulses are discussed in Sec. \ref{SecFinitePulse}).
Considering the evolution from $t=0$ to time $t_F=t_k+\tau_k$ (thus setting  $U(t,t_k+\tau_k)=\mathbb{I}$), one can choose the pulse strength such that the time evolution operator reduces to the squeezing operator that dilates the atomic cloud in coordinate space by a factor $b=b(t_k)$ while compressing it in the momentum representation by a factor $1/b$, 
\beqa
U_{ \delta}(t_F,0)=\mathcal{S}(b)=\exp\left(-i\frac{\log b}{2\hbar}\sum_{i=1}^\N \{\vec{r}_i, \vec{p}_i\}\right).
\eeqa
Indeed,
$\mathcal{S}(b)\Psi(\vec{r}_1,\dots,\vec{r}_\N)=\Psi(\vec{r}_1/b,\dots,\vec{r}_\N/b)$, while for a function in the momentum representation 
$\mathcal{S}(b)\widetilde{\Psi}(\vec{p}_1,\dots,\vec{p}_\N)=\widetilde{\Psi}(\vec{p}_1b,\dots,\vec{p}_\N b)$.

Specifically, this is achieved by choosing the pulse parameters $\tau_k$ and $\om_k$ such that
\beqa
\exp\left[-i\frac{\tau_km\om_k^2}{2\hbar}\sum_{i=1}^{\N} \vec{r}_i\,^{2}\right]\exp\left[i\frac{m\dot{b}}{2\hbar b}\sum_{i=1}^\N \vec{r}_i\,^2\right]=1,\nonumber\\
\eeqa
this is
\beqa
\tau_k\om_k^2=\frac{\dot{b}(t_{k})}{b(t_{k})}.
\label{ekick}
\eeqa
This is the main result of this work and generalizes DKC in the presence of interactions and an arbitrary time-dependent harmonic trap. A further generalization to the case of a nonharmonic trap preserving scale invariance directly follows from \cite{delcampo12,delcampo13,Deffner14}.

\begin{figure}[t]
\includegraphics[width=0.95\linewidth]{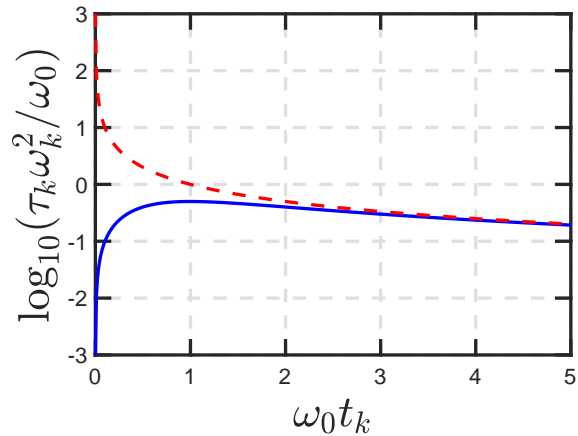}
\caption{\label{Fig1DKC}  Comparison between the pulse parameters determined by  the exact DKC relation  (blue, solid line)  and the long-time limit (red, dashed line).  The kick strength is shown as a function of the free time of flight $t_k$ at which the kick is applied. Differences in the strength between the exact and long-time pulses become significant for $t_k\lesssim 3\om_0^{-1}$.
}
\end{figure}

Under time of flight, the pulse condition (\ref{ekick}) reads
\beqa
\tau_k\om_k^2=\frac{\om_0^2t_k}{1+\om_0^2t_k^2}.
\label{ekickTOF}
\eeqa
For $t_k\gg \om_0^{-1}$, the TOF scaling factor (\ref{btof}) approaches $b(t_k)=\om_0t_k$, and one finds the simplified condition used in the long-time (far-field) DKC \cite{Chu86,Myrskog00} described at the beginning of this section,
\beqa
\label{akickTOF}
\tau_k\om_k^2=1/t_k.
\eeqa
But this is an approximation that is restricted to (i) free TOF under sudden switching of the trap and (ii)  long times of expansion $t_k\gg\om_0^{-1}$.
Figure \ref{Fig1DKC} shows in a TOF setting the difference between the strength of the kick in this limit and the exact result predicted by Eq. (\ref{ekick}).
%

\begin{figure}[t]
\includegraphics[width=0.95\linewidth]{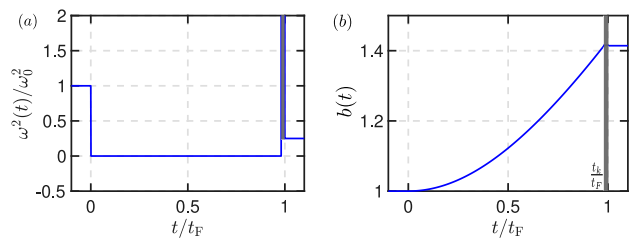}
\caption{DKC after free expansion. (a) Time evolution of $\omega(t)^{2}/\omega^{2}_{0}$ during the process, with parameters $1/5\omega_{k}=2\omega_{F}=\omega_{0}$. (b) Time evolution of $b(t)$. The kick is considered short enough so that evolution of the scaling factor is negligible during its implementation, and the final state is a stationary state of the final trap with frequency $\om_F=\om_0/b_F^2$.
\label{Fig2}}
\end{figure}

 Figure \ref{Fig2} illustrates the associated modulation of the trap frequency and the evolution of the scaling factor during the process. The initial state is in equilibrium with a harmonic trap of frequency $\om_0$. This can be either considered as a reference virtual trap or an initial physical trap  confining the atomic cloud. At $t\geq 0$ the atomic cloud is released and expands freely for a time $t_k$ when a pulse is implemented. Such a pulse can be described by a frequency modulation, although it can be possibly engineered by different means in the laboratory. The atomic cloud right after the pulse is in equilibrium with a trap of frequency $\om_F=\om_0/b_F^2$ with $b_F=b(t_k)$.

\begin{figure}[t]
\includegraphics[width=0.95\linewidth]{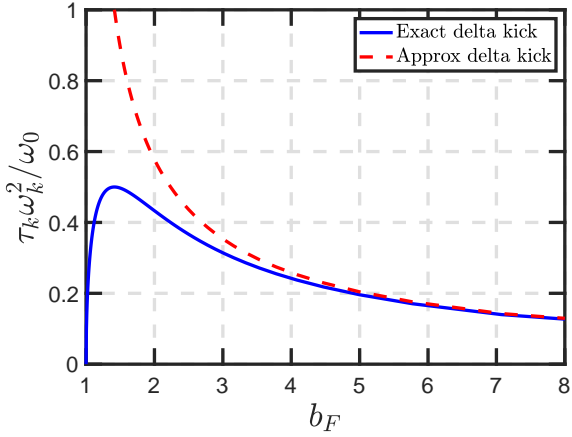}
\caption{Pulse parameters for DKC with free expansion as a function of the final value of the scaling factor, with $b(0)=1$. The blue, solid line corresponds to the exact expression for the kick (\ref{ekickTOF}) and the red, dashed line to the approximated expression (\ref{akickTOF}).
\label{Fig2PPfree}}
\end{figure}

Equation (\ref{ekickTOF}) can be rewritten in terms of the final scaling factor after TOF as $\tau_k\om_k^2/\om_0=\sqrt{b_F^2-1}/b_F^2$.
By contrast, in the long-time approximation $\tau_k\om_k^2/\om_0 =1/\sqrt{b_F^2-1}\approx 1/b_F$.
The agreement between the exact pulse parameters determined by these two approaches is naturally fulfilled whenever the scaling factor is moderately large, as shown in Fig. \ref{Fig2PPfree}.

\section{Phase-space analysis of generalized DKC}

A phase-space description of DKC, generalized for arbitrary modulations of the trapping frequency, makes its operation particularly transparent. The non-commutativity of position and momentum operators prompts the use of the Wigner function, as a quasi-probability distribution \cite{Wigner32,Hillery84}. 
Under scale invariance, using (\ref{psit}) we first note that the one-body reduced density matrix  at time $t$, $\rho_1(t)=\rho_1(\vec{r},\vec{r}\,',t)$, is related to the initial one according to \cite{delcampo11}
\beqa
\rho_1(t)&=&\N\int d\vec{r}_2\cdots \vec{r}_\N\Psi\left(\vec{r}, \vec{r}_2\cdots \vec{r}_\N;t\right)\Psi\left(\vec{r}\,', \vec{r}_2\cdots \vec{r}_\N;t\right)^*\nonumber\\
&=&\frac{\N}{b^D}\exp\left[i\frac{m\dot{b}}{2\hbar b}\sum_{i=1}^\N (\vec{r}\,^2-\vec{r}\,'\,^2)\right]\rho_1\left(\frac{\vec{r}}{b},\frac{\vec{r}\,'}{b},t=0\right).\nonumber\\
\label{rhot}
\eeqa
The Wigner function associated with  the one-body reduced density matrix can then be represented  as a function of the coordinate $\vec{r}$ and the canonically-conjugated momentum  $\vec{p}$,
\beqa
W_t(\vec{r},\vec{p})= \frac{1}{(\pi\hbar)^D} \int \left \langle  \vec{r} -\vec{y}\bigg| \,\hat{\rho}_t  \,\bigg| \vec{r} + \vec{y}  \right\rangle e^{2i \vec{p}\cdot\vec{y}/\hbar}  d \vec{y}\, ,\nonumber
\eeqa
where $\rho(\vec{r},\vec{r}\,')=\langle  \vec{r} | \hat{\rho}_t  | \vec{r}\,'  \rangle$ denotes a density matrix in the coordinate representation. 
The marginals of $W_t(\vec{r},\vec{p})$ correspond to the probability density in the coordinate and momentum representation
\beqa
\int W_t(\vec{r},\vec{p})d\vec{p}&=&\rho(\vec{r},\vec{r})=\langle  \vec{r} | \hat{\rho}_t  | \vec{r}  \rangle,\\
\int W_t(\vec{r},\vec{p})d\vec{r}&=&\rho(\vec{p},\vec{p})=\langle \vec{p} | \hat{\rho}_t  | \vec{p}  \rangle.
\eeqa
 For the description of DKC, it follows from (\ref{rhot}) that the evolution of the Wigner function in a scale-invariant process is governed by a canonical transformation, 
\begin{eqnarray}
\begin{pmatrix}
\vec{r} \\
\vec{p}
\end{pmatrix}
=
\begin{pmatrix}
\alpha & \beta \\
\gamma & \delta
\end{pmatrix}
\begin{pmatrix}
\vec{r}\,' \\
\vec{p}\,'
\end{pmatrix},
\end{eqnarray}
belonging to the two-dimensional real symplectic group ${\rm Sp}(2,\mathbb{R})$.
 The phase-space propagator that determines the evolution of the Wigner function \cite{GarciaCalderon80}
\beqa
W(\vec{r},\vec{p};t)=\iint d\vec{r}\,'d\vec{p}\,'K(\vec{r},\vec{p}|\vec{r}\,',\vec{p}\,')W(\vec{r}\,',\vec{p}\,';0)
\eeqa
becomes
\beqa \label{K_HO}
K(\vec{r},\vec{p}|\vec{r}\,',\vec{p}\,')=\delta[\vec{r}\,'-(\alpha \vec{r}+\beta \vec{p})]\delta[\vec{p}\,'-(\gamma \vec{r}+\delta \vec{p})]\,,\nonumber\\
\eeqa
in terms of the $D$-dimensional delta function.
Under scale invariance (SI), the evolution of an eigenstate at  $t=0$ following a modulation of the trapping frequency  $\om(t)$ is described by \cite{Shanahan18}
\beqa
\begin{pmatrix}
\alpha & \beta \\
\gamma & \delta
\end{pmatrix}_{\rm SI}=
\begin{pmatrix}
1/b & 0 \\
-m\dot{b} & b
\end{pmatrix}\, 
\eeqa
 and the time-dependent Wigner function reads 
\beqa
\label{WtSI}
 W_t(\vec{r},\vec{p})=W_0\left(\frac{\vec{r}}{b},b\vec{p}-m\dot{b}\vec{r}\right),
 \eeqa
 where we note that $W_0$ need not be positive, i.e., it can describe a non-classical state.
 
 The evolution of the Wigner function reflects the fact that as the expansion occurs the momentum distribution is compressed. In addition, the expansion also leads to a momentum shift controlled by the rate of change of the scaling factor.
Both  DKC cooling and STA  aim at engineering an expansion  (or compression) such that the initial Wigner function  $W_0(\vec{r},\vec{p})$ evolves into $ W_t(\vec{r},\vec{p})= W_0(\vec{r}/b,b\vec{p})$, cancelling the momentum shift. While conventional STA  achieve so by reverse engineering the dynamics, the evolution under DKC leads first to the  state (\ref{WtSI}) and then applies a $\delta$-kick (DK),  with phase-space propagator characterized by 
\beqa
\begin{pmatrix}
\alpha & \beta \\
\gamma & \delta
\end{pmatrix}_{\rm DK}=
\begin{pmatrix}
1 & 0 \\
+mb\dot{b} & 1
\end{pmatrix}\, ,
\eeqa
which implements the momentum boost $b\vec{p}-m\dot{b}\vec{r}\rightarrow b\vec{p}$, slowing down the motion of the atomic cloud.
Indeed, the propagator of the complete sequence associated with DKC is set by (\ref{K_HO}) with
\beqa
\begin{pmatrix}
\alpha & \beta \\
\gamma & \delta
\end{pmatrix}_{\rm DKC}=
\begin{pmatrix}
\alpha & \beta \\
\gamma & \delta
\end{pmatrix}_{\rm DK}
\begin{pmatrix}
\alpha & \beta \\
\gamma & \delta
\end{pmatrix}_{\rm SI}=
\begin{pmatrix}
1/b & 0 \\
0 & b
\end{pmatrix},
\eeqa
which guarantees reaching the target state, e.g., with squeezed momentum distribution by a scaling factor $b$.

In short, the condition (\ref{ekick}) specifies the required kick to cancel excitations during an arbitrary scale-invariant expansion, with respect to a Hamiltonian $H(t_F)$ with frequency $\om_F$ in (\ref{omF}).
 This suggests two different strategies to improve DKC: (i) Combine control techniques to engineer expansions faster than TOF and apply modified $\delta$-kicks. (ii) Keeping a trap  turned on  at all times and reverse engineer its frequency $\om(t)$ to prepare the desired target state.
Said differently, one can conceive a new kind of STA: drive $\om(t)$, cancel excitations with a $\delta$-kick and trap the kicked state with a harmonic potential of frequency $\omega_F=\om_0/b^2$.
Effectively, this approach is described by a combination of a smooth driving $\om(t)$ with a  bang-bang protocol with two steps, from $\om(t_k)$ to $\om_k$ and from $\om_k$ to $\omega_F=\om_0/b^2$. 
We discuss these two alternatives below.
Before doing so, we first establish and quantify the advantage of using DKC over adiabatic protocols.

\section{Advantage of DKC with free expansion over adiabatic protocols}
A natural strategy to meet the goal of DKC is to rely on adiabatic driving. Under scale invariance dynamics, the adiabaticity condition was identified by Lewis and Riesenfeld \cite{LewisRiesenfeld69} as
\beqa
\frac{\dot{\om}(t)}{\om(t)^2}\ll 1.
\eeqa
Provided that this condition is satisfied one can engineer an adiabatic expansion to compress the momentum distribution.

\begin{figure}[t]
\includegraphics[width=0.75\linewidth]{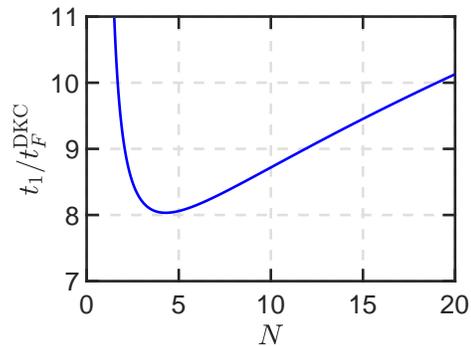}
\caption{Time gain in DKC with free expansion over adiabatic protocols as a function of the the ratio $N=\om_0/\om_F$.
\label{figDKCadiab}
}
\end{figure}

To shorten the process, one may consider protocols in which the nonadiabaticity coefficient is finite, but kept constant at all times.
A protocol satisfying the condition
\beqa
\frac{\dot{\om}(t)}{\om(t)^2}=\mu,
\eeqa
is provided by a time-dependent trap frequency of the form
\beqa
\om(t)=\frac{\om_0t_F}{t_F+(\frac{\om_0}{\om_F}-1)t}, 
\eeqa
that varies from $\om_0$ to $\om_F$ in a time $t_F$. 
In such a protocol 
\beqa
\mu=\left(1-\frac{\om_F}{\om_0}\right)\frac{1}{\om_Ft_F}.
\eeqa
This driving scheme has been explored in a number of works \cite{Rezek09,Uzdin13EP,Beau16}. The evolution of the scaling factor is then non-monotonic and at specific times $\{t_n\}$, 
the rate of change of the scaling factor vanishes, i.e.,  $\dot{b}(t_n) =0$. At these specific times, the scaling factor takes the value
\beqa
b(t_n)=\sqrt{\frac{\om_0}{\om_F}},
\eeqa
which is precisely the result in a truly adiabatic evolution. In this sense, protocols with finite $\mu$ reproduce the exact adiabatic dynamics at the specific instants of time $\{t_n\}$.
It is then possible to stop the expansion at $t_n$ by suddenly changing the trapping frequency from $\om(t_n)$ to $\om_F$, without inducing residual excitations in the final state, which is a stationary state of the final Hamiltonian with frequency $\om_F$.
The shortest time at which this is possible  is  given by \cite{Beau16}
\beqa
 t_1=\frac{1-\frac{\om_F}{\om_0}}{\om_F}\sqrt{1+\frac{4\pi^2}{\ln^2\left(\frac{\om_F}{\om_0}\right)}}.
\eeqa
We note that while the prefactor of the square root is approximately $N/\om_0$ for $N=\om_0/\om_F\gg 1$, the square root term cannot be ignored in the experimentally relevant range of values, e.g., $N=1-100$.
By contrast, relying on DKC with free expansion, the required time for the scaling factor to reach the value $\sqrt{\frac{\om_0}{\om_F}}$ is 
\beqa
t_F^{\rm DKC}=\frac{1}{\om_0}\sqrt{\frac{\om_0}{\om_F}-1}\approx \sqrt{N}/\om_0.
\eeqa
The ratio of the required times is thus  
\beqa
\label{tratio}
\frac{t_1}{t_F^{\rm DKC}}=\sqrt{(N-1)\left(1+\frac{4\pi^2}{\ln^2\left(N\right)}\right)},
\eeqa
which as a function of $N$ takes the minimum value of $8.03$ for $N=4.29$. This provides the minimum time gain by DKC over adiabatic approaches. Figure \ref{figDKCadiab} shows the actual time gain as a function of $N$. While asymptotically it approaches $\sqrt{N}$,  the actual expression (\ref{tratio}) is required to estimate the advantage of DKC over a (quasi-) adiabatic protocol for realistic values of $N$.

\section{Sudden trap-frequency quenches and time-optimal control \label{sec:opti}}
Consider an initial atomic cloud confined in a trap of frequency $\om_0$ and released into a weaker trap of frequency $\om_1$.
Provided the dynamics is scale invariant, the time evolution of the radius of the atomic cloud is governed by the scaling factor. The solution of the Ermakov equation can generally be determined as detailed in the Appendix \ref{app:solerma}  and in this case is well known \cite{Pinney50,MinguzziGangardt05}
\beqa
b(t)=\sqrt{1+\left(\frac{\om_0^2}{\om_1^2}-1\right)\sin^2(\om_1 t)}.
\eeqa
This solution is oscillatory between the minimum value $b_{\rm min}=1$ and the maximum value $b_{\rm max}=\frac{\om_0}{\om_1}$ with frequency $2\om_1$ and period $T=\pi/\om_1$.
We note that $b_{\rm max}=b_{\rm ad}^2$, where $b_{\rm ad}=\sqrt{\frac{\om_0}{\om_1}}$ is the solution obtained in the adiabatic limit by setting $\ddot{b}\approx 0$ in the Ermakov equation.
Specifically, we note that at 
\beqa
t_n=n\frac{\pi}{2\om_1},\quad b(t_n)=\frac{\om_0}{\om_1}, \quad \dot{b}(t_n)=0,
\eeqa
and the time evolving eigenstate becomes a stationary eigenstate of a trap with frequency $\om_1^2/\om_0$.
It is thus possibly to suddenly switch at $t=t_n$ between the frequency $\om_1$ and $\om_F=\om_1^2/\om_0$ to prepare a stationary state in the final trap with frequency $\om_1^2/\om_0$.
The time-optimal sequence involves two sudden quenches of the trapping frequency, one at $t=0$ from $\om_0$ to $\om_1$, and a second sudden quench at $t_1=\frac{\pi}{2\om_1}$ from $\om_1$ to $\om_1^2/\om_0$. This is in essence a bang-bang protocol.
And it sets the standards: to beat this protocol one needs to prepare a stationary state with final frequency $\om_1^2/\om_0$ in a time shorter than  $t_1=\frac{\pi}{2\om_1}$.
As it turns out, this is always possible using STA provided that an arbitrary $\om(t)$ can be implemented \cite{Chen10,delcampo11}. 

It is instructive to compare such sudden quench with DKC using free flight. Considering a thermal state of a harmonic oscillator with initial inverse temperature $\beta_0$ and frequency $\om_0$, phase-space preserving cooling allows one  to prepare final states with frequency $\om_F$ and inverse temperature $\om_F$, provided that $\beta_0\om_0=\beta_F\om_F$. To reduce the temperature by a factor $N$ such that $T_F=T_0/N$ one must target a frequency $\om_F=(T_F/T_0)\om_0=\om_0/N$, e.g., under adiabatic driving.
Using the above bang-bang protocols with $\om_1=\sqrt{\om_0\om_F}$ one can succeed in a time 
\beqa 
t_1=\frac{\pi\sqrt{N}}{2\om_0}.
\eeqa

Said differently, for a final $\om_F=\om_0/N$ using $\om_F=\om_0/b^2$ and $b(t_n)=\frac{\om_0}{\om_1}$, one finds $\om_F=\om_1^2/\om_0$, i.e., $\om_1=\sqrt{\om_0\om_F}$. This is achieved in a time 
\beqa
t_1=\frac{\pi}{2\sqrt{\om_0\om_F}}=\frac{\pi\sqrt{N}}{2\om_0}
\eeqa
 while standard DKC  requires 
\beqa
\om_F=\frac{\om_0}{1+\om_0^2t^2}=\frac{\om_0}{N},
\eeqa
and thus a time 
\beqa
t&=&
\label{tDKCTOF}
\frac{1}{\om_0}\sqrt{\frac{\om_0}{\om_F}-1}\\
&\approx&\frac{\sqrt{N}}{\om_0}.
\eeqa 
As expected, the two protocols are approximately equivalent given  the time scales involved. However, there is no need to perform a kick when releasing the cloud in a weaker trap, as the later can be used to slow down the particles with an effect equal to that intended by a perfect kick. Conversely, DKC replaces the need for a  final trap by a pulse.

What is the relation of these protocols to optimal control theory?
The time-optimal bang-bang protocol whenever the trap frequency is restricted to be positive $\om(t)\geq 0$ and within the interval  
$\om_{\rm min}\leq \om(t)\leq \om_{max}$ has been provided by Salamon et al. for an  intermediate frequency change \cite{Salamon09}

\begin{equation}
\omega(t)=
\begin{cases}
\omega_{0} &  t \leq 0 \\
\omega_{1} & 0 < t <t_1\\
\omega_{2} & t_1 <t < t_1 + t_2\\
\omega_{F} & t \geq  t_F=t_1 + t_2.
\end{cases}
\end{equation}

In the limit, $\om_1\rightarrow0$, $\om_2\rightarrow\infty$, one finds the fastest  protocol
with a total expansion time (\ref{tDKCTOF}).
Indeed, when $\om_1\rightarrow0$ (free expansion), $b(t)=\sqrt{1+\om_0^2t^2}$ for $0\leq t\leq t_1$. 
Identifying $t_1=t_k$, the $\om_2\rightarrow\infty$ can be implemented by a $\delta$-kick, as those used in DKC, making $\tau_k=t_2\rightarrow0$.
The total expansion time is then governed by the required time for the scaling factor under free flight to match the equilibrium value corresponding to the target final trap of frequency $\om_F$,
\beqa
\sqrt{1+\om_0^2t^2}= \sqrt{\frac{\om_0}{\om_F}},
\eeqa
that 
yields as a solution Eq. (\ref{tDKCTOF}). 

 As a result,
DKC with free expansion 
is equivalent to the time-optimal control with unbounded trapping frequencies. Said differently, as long as the trap frequency takes only real values during the expansion within the range $[\om_{\rm min},\om_{\rm max}]$, DKC with free expansion provides the time-optimal protocol (provided that a $\delta$-kick can be implemented). As we next show, the relation between DKC and time-optimal control can as well be extended to protocols involving trap inversion.

\section{DKC with trap inversion and time-optimal control \label{sec:bgbg}}

Given an initial equilibrium state confined in a trap of frequency $\om_0$, consider the sudden inversion of the trap, by making the frequency purely imaginary $\om_0\longrightarrow \om_1=i\om_I$. 
This amounts to a repulsive rather than an attractive potential, with the profile of an inverted parabola.
Such a quench can cause more rapid expansion than conventional DKC and thus generally leads to faster protocols. To appreciate this, consider the trajectory of a single particle with the equations of motion $\dot{\vec{r}}=\vec{p}/m$ and $\dot{\vec{p}}=+m\om_I^2\vec{r}$. The following  quantity  
\beqa
\dot{\vec{z}}=\frac{d}{dt}\left(\frac{{\vec{p}}}{\sqrt{m}}+\sqrt{m\om_I^2}{\vec{r}}\right)=\om_I\vec{z},
\eeqa
grows exponentially in time as $\vec{z}(t)=\vec{z}(0)\exp{\om_I t}$.
 As a result, given an initial phase-space distribution, the nearly exponential spreading in coordinate space yields a nearly-exponential squeezing of the momentum distribution, given Liouville's theorem. In what follows, we provide an exact account in terms of the scale-invariant dynamics.

\begin{figure}[t]
\includegraphics[width=0.95\linewidth]{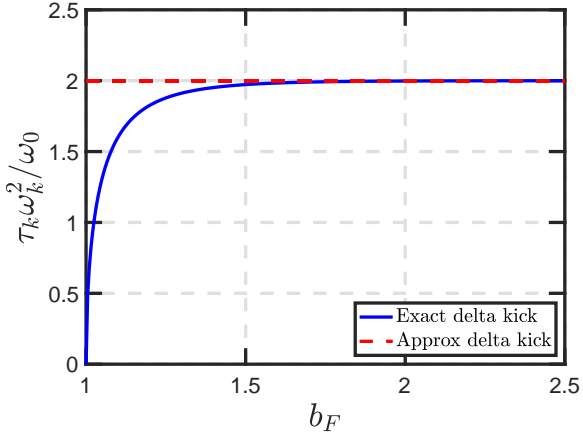}
\caption{Pulse parameters for DKC with trap inversion as a function of the final value of the scaling factor. The blue, solid line corresponds to the exact expression for the kick (\ref{eq:tauTIex}) and the red, dashed line to the approximated expression (\ref{eq:tauTIapp}).
\label{FigPPTI}}
\end{figure}

\begin{figure}[t]
\includegraphics[width=0.95\linewidth]{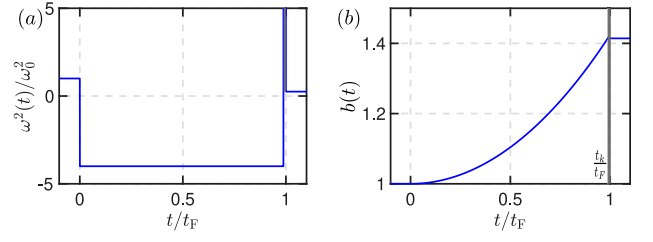}
\caption{DKC with trap inversion. (a) The trap inversion corresponds to a sudden quench of the frequency-square   from its initial value $\omega^{2}_{0}$ to a negative one $-\omega^{2}_{I}$, yielding a nearly exponential expansion. At time $t_k$, a kick of frequency $\omega_{k}$ (vertical gray line) is applied, bringing the atomic cloud at equilibrium with the final trap  of frequency $\omega_{F}$.  (b) The scaling factor approaches an exponential expansion, that is terminated by the kick. In the instantaneous pulse approaches, the evolution of the scaling factor is considered negligible during the pulse and remains at equilibrium afterwards. Here, $\omega_{0}=1, \omega^{2}_{k}=200\omega^{2}_{0}$, and $\omega_{I}/4=2\omega_{F}=\omega_{0}$.
\label{FigPPTIprot}}
\end{figure}

The scaling factor following the quench grows as \cite{Chen10}
\beqa
\label{bexp}
b(t)&=&\sqrt{1+\left(\frac{\om_0^2}{\om_I^2}+1\right)\sinh^2(\om_I t)}\nonumber\\
&\approx&\frac{1}{2}\left(\frac{\om_0^2}{\om_I^2}+1\right)^{\frac{1}{2}}\exp(\om_It), 
\eeqa
where the approximation in the second line holds for $t\gg\om_I^{-1}$, i.e., in the long-time limit.
The resulting non-equilibrium state has the same width as that of an equilibrium state  of the Hamiltonian $H(t)$ in  a trap of frequency $\om(t)=\om_0/b(t)^2$ and excitations can be cancelled with a $\delta$-kick at $t=t_k$ satisfying
\beqa
\tau_k\om_k^2&=&\frac{\dot{b}}{b}=\frac{\om_I}{2}\frac{(\om_0^2+\om_I^2)\sinh(2\om_I t_k)}{\om_I^2+(\om_0^2+\om_I^2)\sinh^2(\om_I t_k)}
\label{eq:tauTIex}\\
&\approx&\om_I.
\label{eq:tauTIapp}
\eeqa 
 The second line follows from the approximate exponential growth of the scaling factor for $t_k\gg\om_I^{-1}$, which makes the kick robust against errors in the timing of $t_k$.
As a function of the final value of the expansion factor, the comparison between (\ref{eq:tauTIex}) and (\ref{eq:tauTIapp}) is shown in Fig. \ref{FigPPTI}, with an excellent agreement being found for $b_F>3/2$.
The modulation of the trapping frequency for an example protocol is shown in Fig. \ref{FigPPTIprot}, together with the evolution of the scaling factor.

\begin{figure}[t]
\includegraphics[width=0.95\linewidth]{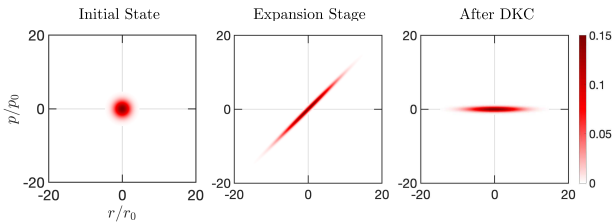}
\caption{Phase space diagrams of DKC with trap inversion with the same expansion time as in Fig. \ref{phase_space} (i.e. $t_F=3/2\omega_0$) and with $\omega_I=\omega_0$. Note that DKC with trap inversion increases the momentum spread during expansion and thus requires a  stronger kick.\label{phase_spaceTI}}
\end{figure}

Let us compare the free TOF used in the standard DKC (Fig. \ref{phase_space}) with the expansion under trap inversion (TI) (Fig. \ref{phase_spaceTI}), fixing the expansion time in both cases equal to $t_F$. Specifically, considering the ratio that governs the gain in the squeezing of the momentum distribution
\beqa
\frac{\Delta  R(t_F)}{[\Delta R(t_F)]_{\rm TI}}\approx \frac{1}{2}\left(\frac{1}{\om_0^2}+\frac{1}{\om_I^2}\right)^{\frac{1}{2}}\frac{\exp(\om_It_F)}{t_F},
\eeqa
one finds that  the  degree of momentum squeezing under trap inversion greatly exceeds that achievable by standard DKC. 
However, it should be taken into account that the strength of the pulse that is to be applied in DKC with trap inversion is higher than under free expansion. Indeed, the ratio between the pulse parameters in both cases is set by the ratio between (\ref{ekickTOF}) and (\ref{eq:tauTIex}), that approximately reduces to
\beqa
\frac{\tau_k\om_k^2}{[\tau_k\om_k^2]_{\rm TI}}\approx \frac{1}{\om_I t_k}.
\eeqa
In the laboratory, only pulses with a maximum value of $\tau_k\om_k^2$ can be implemented, usually because of restrictions on the attainable trap frequency in conjunction with the nonharmonic shape of most traps far from their center. Hence, DKC with trap inversion may impose more demanding practical requirements.

Alternatively, the comparison between different protocols can be done  by considering the required time for the scaling factor to reach  a fixed, target value  $b(t_F)=b_F$.
After a sudden trap inversion, the required expansion time is
\beqa
\label{toptinv}
t_F&=&\frac{1}{\om_I}\sinh^{-1}\sqrt{\frac{b_F^2-1}{\left(\frac{\om_0^2}{\om_I^2}+1\right)}}\\
&\approx&\frac{1}{\om_I}\log\left(\frac{2b_F}{\left(\frac{\om_0^2}{\om_I^2}+1\right)^{\frac{1}{2}}}\right).
\eeqa
By contrast, DKC relying on  free TOF expansion requires a time $t_F^{\rm DKC}\approx b_F/\om_0$ and thus
\beqa
\frac{t_F}{t_F^{\rm TI}}\approx \frac{\om_0}{\om_I b_F}\log\left(\frac{2b_F}{\left(\frac{\om_0^2}{\om_I^2}+1\right)^{\frac{1}{2}}}\right),
\eeqa
providing a great speedup as a function of $b_F$ and $\om_0/\om_I$.

It is worth establishing the relation between such protocol and the time-optimal bang-bang protocol allowing for trap inversion, that is, with a purely imaginary frequency.
In the context of STA, Chen et al.   \cite{Chen10} proposed a bang-bang protocol involving a sudden inversion of the trap frequency for a given time $t_1$, with $\om(t)=i\om_I$
and a subsequent sudden quench to a trap of frequency $\om_2$ with a waiting time $t_2$.
DKC with trap inversion replaces the second stage of the STA expansion during the time $t_2$ by a $\delta$-kick pulse of vanishing duration.
As a result,  DKC with trap inversion shortens the expansion time with respect to the protocol proposed in \cite{Chen10}, achieving the target state 
in an expansion time that is solely given by $t_1=t_k$ in Eq.  (\ref{toptinv}).
Chen et al. made however no claim on time-optimality in the use of an inverted harmonic trap. 
The time-optimal bang-bang protocol with trap inversion was reported in \cite{Stefanatos10}.
In the case of unbounded frequency $\om_2$, and with $\om_I\geq \om_0$, the time-optimal protocol was shown to take the form in Eq.  (\ref{toptinv}), which recovers Eq. (36) in \cite{Stefanatos10}.
As long as the duration of the $\delta$-kick pulse is negligible, the protocol involving a sudden trap inversion and a $\delta$-kick is thus time-optimal.

\section{Shortcuts to adiabaticity assisted by $\delta$-kicks}

In what follows, we consider protocols by easing the requirement for time-optimality. As we shall see, this allows us to introduce a new class of STA combining  $\delta$-kicks with the modulation of the trap-frequency based on reverse engineering the scale-invariant dynamics. We term such protocols $\delta$-STA for short. This new class of STA provides a higher level of freedom in the design of the control protocol than that offered by techniques previously reported in the literature \cite{Salamon09,Chen10,Muga09,Stefanatos10,Hoffmann11,delcampo11,Choi2011,Choi2011b,delcampo12,Jarzynski13,delcampo13,Deffner14}. This feature is particularly advantageous when the trapping confinement and pulse implementation are achieved by different means, as it is often the case in the laboratory. 

Consider the control task of driving an initial stationary state of $H(0)$ with frequency $\om_0$ to a target state of a Hamiltonian  $H(t_F)$ with frequency $\om_F$ in a prescheduled time $t_F$.
We choose $\om_F=\om_0/b_F^2$ and reverse engineer the scaling factor $b(t)$ so that it matches the usual boundary conditions for an initial stationary state
\beqa
b(0)=1, \dot{b}(0)=0,
\eeqa
which make the time-dependent state $\Psi(t)$ equal to $\Psi(0)$ at $t=0$. 

Regarding the evolution of the state at a later time, we simply consider
\beqa
b(t_k)=b_F=\sqrt{\frac{\om_0}{\om_F}},
\eeqa
with no condition on the derivatives $\dot{b}(t_k)$ and $\ddot{b}(t_k)$.
A polynomial interpolating ansatz satisfying  all these conditions is
\beqa
\label{bn1}
b(t)=1+(b_F-1)\left(\frac{t}{t_k}\right)^{2}.
\eeqa
The associated control protocol consists of three steps.
First, one implements the following driving  during the time interval $t\in[0,t_k)$.
\beqa
\om(t)^2=\frac{\om_0^2}{b^{4}}- \frac{\ddot{b}}{b}=\frac{\om_0^2}{b^{4}}-\frac{2(b_F-1)}{t_k^2b},
\eeqa
where we have used the Ermakov equation.
Note that $\om(t_k)\neq\om_F$, away from the adiabatic limit. Indeed, at the end of the interval $\om(t_k)^2=\om_F^2-\frac{2(b_F-1)}{t_k^2b_F}$.

Subsequently, at $t=t_k$,  a $\delta$-kick is applied, fulfilling the condition 
\beqa
\tau_k\om_k^2=\frac{\dot{b}(t_k)}{b(t_k)}=\frac{2(b_F-1)}{t_k b_F}.
\eeqa

Finally, at $t_F=t_k+\tau_k$ a trap with frequency 
\beqa
\label{omFeq}
\om_F=\frac{\om_0}{b_F^2}
\eeqa
is turned on.

\begin{figure}[t]
\includegraphics[width=1\linewidth]{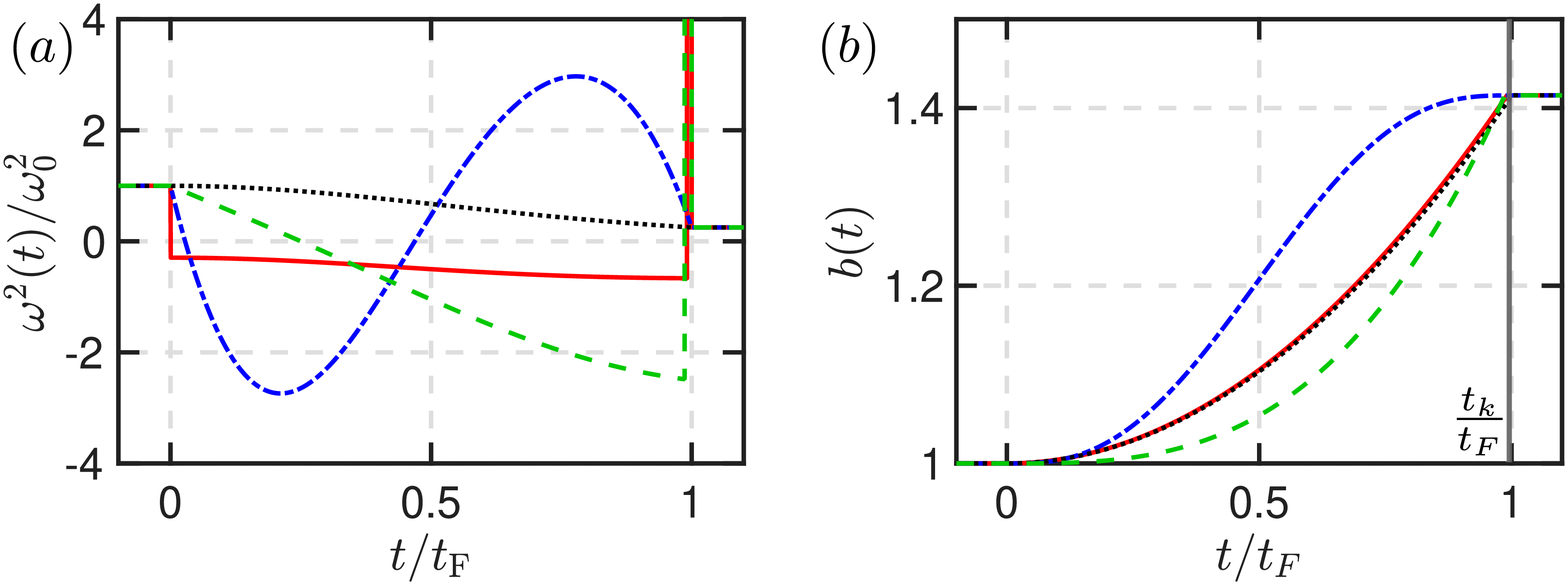}
\caption{Comparison of $\delta$-STA and STA based on reverse-engineering. (a) Modulation of the trapping frequency in  $\delta$-STA (red solid line) and STA by reverse engineering (dotted-dashed blue line) \cite{Chen10}. An ``adiabatic'' trajectory $\om_0^2/b(t)^4$ with $b(t)$ in Eq. (\ref{bn1}) is shown as a reference (dotted black line). A $\delta$-STA involves a sudden quench at $t=0$ and a pulse before the end of the protocol, with a rather gentle modulation of the trap frequency in between. A conventional STA involves high transient frequencies in the intermediate stages, instead of a pulse. 
Suppressing the sudden quench at $t=0$ in $\delta$-STA with $n=2$ (dashed green line) leads to higher values of $\om_I$, still below those involved in conventional STA.
(b) Evolution of the scaling factor, showing that a $\delta$-STA  achieves higher expansion rates than  a conventional STA during the protocol. 
While in conventional STA the process of slowing down extends during the second half of the protocol, it is sharply localized at $t_k$ in a $\delta$-STA  (vertical gray line). 
}\label{fig:STAvsSTA}
\end{figure}

$\delta$-STA offer advantages with respect to STA designed by reverse engineering, e.g., \cite{Chen10,delcampo11}. In particular, by combining sudden quenches and $\delta$-kicks they make possible to engineer driving protocols between the same initial and final states that lower the required values of the transient interpolating driving frequency $\om(t)$, as shown in Fig. \ref{fig:STAvsSTA}.

%

In short, the above protocol involves a sudden quench from $\om_0$ to $\om(0)$, the modulation $\om(t)$, a $\delta$-kick, and a final sudden quench. One can remove the first sudden quench by imposing $\om(t=0)=\om_0$, which is satisfied if $b(0)=1$ together with  the auxiliary condition $\ddot{b}(0)=0$. Indeed, one can readily consider  supplementary boundary conditions associated with  higher-order vanishing derivatives
\beqa
\ddot{b}(0)=\dddot{b}(0)=\cdots = b^{n)}(0)=0,
\eeqa
where $b^{n)}$ denotes the $n$-th time derivative.
These are  satisfied by an interpolating polynomial
\beqa
b(t)=1+(b_F-1)\left(\frac{t}{t_k}\right)^{n+1}.
\eeqa
that results from a trap frequency modulation governed by the Ermakov equation, $\om(t)^2=\frac{\om_0^2}{b^{4}}- \frac{\ddot{b}}{b}$, reaching right before the kick the value  $\om(t_k)^2=\om_0^2/b_F^4-n(n+1)(b_F-1)b_F^3/t_F^2$.
For $n\geq 2$, $\om(t)=\om_0$, thus removing the need for a sudden quench at $t=0$.  As for the pulse to be applied at $t=t_k$, it is characterized by
\beqa
\tau_k\om_k^2=\frac{\dot{b}(t_k)}{b(t_k)}=\frac{(n+1)(b_F-1)}{t_k b_F},
\eeqa
 which increases with $n$, potentially precluding its implementation in the laboratory if a maximum value of $\tau_k\om_k^2$ is at reach in a given setup.
Upon completion of the protocol, the final state  is the same as with $n=1$ and is a stationary state of a  trap of  frequency $\om_F=\om_0/b_F^2$.

\section{DKC with pulses of finite duration and strength}\label{SecFinitePulse}

\subsection{DKC as a limit of time-optimal control \label{sec:finite}}
So far we have consider DKC  associated with an idealized instantaneous $\delta$-kick  (\ref{eq:delta}). However, the implementation in  the laboratory of such pulses generally takes a finite amount of time and makes use of a finite potential.
To explore their effect, we consider the canonical DKC under free time-of flight as a bang-bang control protocol.
In this case, the  width of the atomic cloud matches a harmonic trap of frequency $\om_0$ (virtual or real) that is switched off at $t=0$. Free evolution ensues for a time $t_k$ and is interrupted by a pulse of duration $\tau_k$ associated with a trap of finite frequency. 
We thus consider a finite-frequency bang-bang control protocol
\begin{equation}
\label{bbom2}
\omega(t)=
\begin{cases}
\omega_{0} &  t \leq 0 \\
0 & 0 < t <t_k\\
\omega_{k} & t_k <t < t_k + \tau_k\\
\omega_{F} & t \geq  t_F=t_k + \tau_k.
\end{cases}
\end{equation}

The evolution of the scaling factor  during the full process is determined by the Ermakov equation (\ref{eq:ermakov}) and reads  \cite{Stefanatos10}
\begin{eqnarray}
b(t)=
\begin{cases}
b_{\rm TOF}(t) & 0 < t < t_k,\\
b_{2}(t) &  t_k < t < t_{F}.
\end{cases}
\end{eqnarray}
with $b_{\rm TOF}(t)=\sqrt{1+(\omega_{0}t)^{2}}$ and 
\beqa
b_{2}(t)=\sqrt{b^{2}_{F}+\left[\left(\frac{\omega_{0}}{b_{F}\omega_{k}}\right)^{2}-b^{2}_{F}\right]\sin^{2}[\omega_{k}(t-t_F)]},\nonumber\\
\eeqa
with $b_F=\sqrt{\om_0/\om_F}$. 

We note that $b_2(t)$ is found by reversing the dynamics from the target state at the end of the process $t_F$.
As detailed in Appendix \ref{app:duration},  imposing the continuity condition of the scaling factor at the time of the kick,  $b_{\rm TOF}(t_{k})=b_{2}(t_{k})$, determines both the  kick time $t_{k}$ and the kick duration $\tau_k$, that are set by
\begin{eqnarray}
\label{eq:t1t2}
t_k&=&\frac{1}{\omega_{0}}\sqrt{b^{2}_{F}-1+\frac{1-b^{2}_{F}}{b^{2}_{F}}\left(\frac{\omega_{0}}{\omega_{k}}\right)^{2}},\\
\label{etaukfp}
\tau_k&=&\frac{1}{\omega_{k}}{\rm arcsin}\sqrt{\frac{b^{2}_{F}-1}{b^{4}_{F}\frac{\omega^{2}_{k}}{\omega^{2}_{0}}-1}}.
\end{eqnarray}
Importantly, with access to unbounded frequencies, in the limit $\omega_{k}\to + \infty $, $\tau_k$ vanishes and one recovers   $t_{k}\rightarrow t_{F}=\frac{\sqrt{\omega_{0}/\omega_{F}-1}}{\omega_{0}}$, in agreement with Eq. (\ref{tDKCTOF}). 
As for the exact finite-pulse duration (\ref{etaukfp}), in this limit
\beqa
\label{bbDKC}
\tau_k\om_k^2=\frac{\sqrt{b_F^2-1}}{b_F^2}\om_0,
\eeqa
which is precisely the value predicted by the exact instantaneous pulse  condition
Eq. (\ref{ekickTOF}), for the chosen $t_k$.
This proves that DKC can be derived as a limit of the bang-bang protocol when the pulse frequency is unbounded. 
In this sense, DKC is time-optimal.

Further, we note that (\ref{bbDKC}) differs from the usual DKC condition 
Eq. (\ref{akickTOF}) derived from the long-time classical dynamics of point-like particles \cite{Chu86}, 
\beqa
\tau_k\om_k^2=1/t_k=\frac{1}{\sqrt{b_F^2-1}}\om_0\approx \frac{\om_0}{b_F}.
\eeqa
\begin{figure}[t]
\includegraphics[scale=0.35]{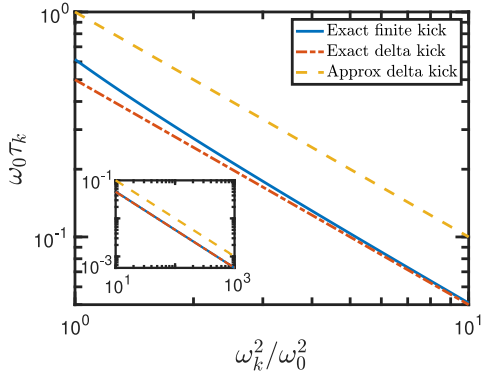}
\caption{DKC with finite pulses. Comparison of the pulse parameters in the exact finite-pulse analysis (\ref{etaukfp}), the exact $\delta$-kick treatment (\ref{ekickTOF}) and long-time $\delta$-kick approximation (\ref{akickTOF}), taking  $\omega_{F}=\omega_{0}/2$. \label{fig:kicktimes}
}
\end{figure}
In Fig. \ref{fig:kicktimes} we compare the pulse parameters given by the different approaches, including the exact treatment with a $\delta$-kick  (\ref{ekickTOF}), the long-time (far-field) classical approximation  for a $\delta$-kick  (\ref{akickTOF}), and the exact finite-pulse condition (\ref{etaukfp}).
The pulse parameters for the exact finite-kick and  $\delta$-kick models agree  for $\omega_{k} \geq 4 \omega_{0}$.  Note that the parameters are such that $b_F=\sqrt{2}$. The validity of the long-time DKC protocol further requires  $b_F\gg1$, so that $\tau_k\om_k^2\approx \om_0/b_F$.


\subsection{Variation of the scaling factor during a finite pulse}

Away from the $\delta$-kick limit, the scaling factor $b(t)$ may evolve during the application of a pulse of finite duration.  Using the finite-pulse expression for scaling factor allows us to quantify this deviation.
We thus consider the modulation in (\ref{bbom2}). 
Instead of imposing boundary conditions at the beginning and end of the process and using a mixed forward-backward interpolation of the scaling factor, as done in the previous section, we use a forward evolution in which the magnitude and rate of the scaling factor
at $t_k$ are used as initial conditions for the subsequent stage, from $t_k$ to $t_F$.
The solution of the Ermakov equation (\ref{eq:ermakov})  during the full process reads 
\begin{eqnarray}
b(t)=
\begin{cases}
b_{\rm TOF}(t) & 0 < t < t_k,\\
b_{2}(t) &  t_k< t < t_{F}.
\end{cases}
\end{eqnarray}
with $b_{\rm TOF}(t)=\sqrt{1+\omega_{0}^2t^{2}}$ while for $t\geq t_k$ the evolution of the scaling factor can be founds from using Pinney's method as detailed in the Appendix \ref{app:solerma} and reads
\beqa
b_2= \!\left[\!\left(b_k\cos(\omega_k s)+\frac{\dot{b}_k\sin(\omega_k s)}{\omega_k}\right)^{2}\!\!\!+
\frac{\omega_{0}^2}{b_k^2\omega_k^2}\sin^{2}(\omega_k s)\right]^{\frac{1}{2}}.\nonumber\\
 \eeqa  
with $s=t-t_k$ and
\beqa
 b_k&=&b_{\rm TOF}(t_{k})=\sqrt{1+\omega_{0}^2t_{k}^{2}}, \\
  \dot{b}_k&=&\dot{b}_{\rm TOF}(t_{k})=\frac{\omega^{2}_{0}t_{k}}{\sqrt{1+\omega_{0}^2t_{k}^{2}}}. 
 \eeqa
 In the spirit of (\ref{tDKCTOF}), one can choose the kick time $t_k$ such that $b_{\rm TOF}(t_{k})=b_F$, which 
determines $t_{k}=\sqrt{b^{2}_{F}-1}/\omega_{0}$ as well as $\dot{b}(t_{k})=\omega_{0}\sqrt{b^{2}_{F}-1}/b_{F}$. 

The exact variation of the scaling factor during the completion of the pulse is given by 
\beqa
\delta b_k=b_2(t_k+\tau_k)-b_k.
\eeqa
In the instantaneous pulse approximation, $\tau_k$ is chosen according to Eq. (\ref{ekickTOF}) -see (\ref{bbDKC})- which yields
\beqa
\tau_k\om_k=\sqrt{1-\frac{1}{b_F^2}}\frac{\om_0}{b_F \om_k}.
\eeqa
This can be used as a small expansion parameter provided that $b_F\om_k\gg \om_0$.
To leading order in $\omega_{k}\tau_{k}$,
\begin{equation}
\delta b_k=\frac{b^{2}_{F}-1}{b_F^3}\left(\frac{\om_0}{\om_k}\right)^2.
\end{equation}
For the instantaneous pulse approximation to hold, one requires $\delta b_k\ll 1$, which is guaranteed provided that $\om_0/\om_k\ll 1$.

Further, upon completion of the pulse, in the instantaneous approximation $\dot{b}_F=\dot{b}(t_k+\tau_k)=0$.
With a finite pulse analysis, one finds to leading order in $\tau_k\om_k$, that
\beqa
\dot{b}_F=\frac{\sqrt{b_F^2-1}}{b_F^5}\om_0\left(\frac{\om_0}{\om_k}\right)^2,
\eeqa
that is negligible, $\dot{b}_F\approx 0$, in the limit $\om_0/\om_k\ll 1$.

\subsection{DKC with trap inversion from optimal control}

We demonstrated that the $\delta$-kick is the limit process of the bang-bang trajectory with positive frequency in Sec. \ref{sec:opti}. As we next show, the protocol introduced in Sec. \ref{sec:bgbg} can be also understood as the limit of a bang-bang process with an  inverted-trap and an intermediate frequency change. The time-optimal protocol whenever the trap frequency can be purely imaginary $\om^{2}(t) < 0$ and within a given frequency range  
$\om_{\rm min}\leq |\om(t)| \leq \om_{max}$ has been found using Pontryagin’s maximum principle by Stefanatos et al.  \cite{Stefanatos10}
\begin{equation}
\omega(t)=
\begin{cases}
\omega_{0} &  t \leq 0 \\
i\omega_{I} & 0 < t <t_k\\
\omega_{k} & t_k <t < t_k + \tau_k\\
\omega_{F} & t \geq  t_F=t_k + \tau_k.
\end{cases}
\end{equation}
The variation of the width of the cloud is governed by the evolution of the scaling factor, which can be readily found by solving the Ermakov equation and takes the form
\begin{eqnarray}
b(t)=
\begin{cases}
b_{1}(t) & 0 < t < t_{k},\\
b_{2}(t) &  t_{k} < t < t_{k}+\tau_{k},
\end{cases}
\end{eqnarray}
where
 \beqa
\label{b1wI}  b_{1}(t)&=&\sqrt{1+\left(1+\left(\frac{\omega_{0}}{\omega_{I}}\right)^{2}\right)\sinh^{2}(\omega_{I} t)}\\
 b_{2}(t)&=&\sqrt{b^{2}_{F}+\left(\left(\frac{\omega_{0}}{b_{F}\omega_{k}}\right)^{2}-b^{2}_{F}\right)\sin^{2}(\omega_{k}(t-t_{F}))}.\nonumber\\
 \eeqa 
 The expansion in the inverted trap occurs  for a time
\begin{eqnarray}
\label{eq:tkwI}
t_{k}=\frac{1}{\omega_{I}}{\rm arcsinh}\sqrt{\frac{v_{1}(b^{2}_{F}-1)(b^{2}_{F}v_{2}-1)}{b^{2}_{F}(v_{1}+v_{2})(v_{1}+1)}}.
\end{eqnarray}
with $v_{1}=\left(\frac{\omega_{I}}{\omega_{0}}\right)^{2},v_{2}=\left(\frac{\omega_{k}}{\omega_{0}}\right)^{2}$, 
as detailed  in Appendix  \ref{app:duration}, see as well \cite{Stefanatos10}.

The subsequent kick with finite frequency $\om_k$ is applied for a time
\begin{equation}
\label{eq:t2}
\tau_k=\frac{1}{\omega_{k}}{\rm arcsin}\sqrt{\frac{v_{2}(b^{2}_{F}-1)(b^{2}_{F}v_{1}+1)}{(v_{1}+v_{2})(b^{4}_{F}v_{2}-1)}},
\end{equation}
which sets the exact relation $\tau_k\om_k^2$ for DKC with an inverted trap with a pulse of finite duration and strength.
In particular, in the limit of an unbounded  pulse frequency $\omega_{k}$,
\begin{equation}
\label{eq:taukwk2wI}
\tau_k\omega^{2}_{k}=\frac{\omega_{0}}{b^{2}_{F}}\sqrt{(b^{2}_{F}-1)\left(b^{2}_{F}\frac{\omega_{I}^2}{\omega_{0}^2}+1\right)}.
\end{equation}

The explicit form of the kick time $t_k$ (\ref{eq:tkwI}) in the time-optimal protocol is not a priori known in DKC with $\delta$-kicks.
Targeting a final expansion factor $b_F$, the expansion with the inverted trap  terminates at  the kick time (\ref{toptinv}). In this case, the DKC condition  (\ref{ekick}), determined using  Eqs. (\ref{b1wI}) and (\ref{toptinv}), reduces precisely to (\ref{eq:taukwk2wI}). This proves the equivalence between DKC and time-optimal bang-bang protocols with access to trap inversion, in the limit of a large pulse frequency.

\begin{figure}[t]
\includegraphics[scale=0.35]{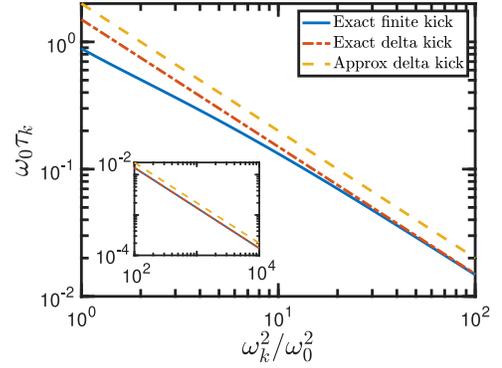}
\caption{DKC with finite pulses and  trap inversion.  The duration of the kick  $\tau_{k}$ is plotted as a function of the pulse frequency, when determined by the  exact finite-pulse relation (\ref{eq:t2}), the exact  $\delta$-kick  (\ref{eq:tauTIex}), and the long-time approximate $\delta$-kick (\ref{eq:tauTIapp}) , with $\omega_{F}=\omega_{I}/4=\omega_{0}/2$. \label{fig:bgbgtau}}
\end{figure}

 In Fig. \ref{fig:bgbgtau} we compare  the exact expression for a finite pulse using (\ref{eq:t2}), the exact expression for an instantaneous $\delta$-kick (\ref{eq:tauTIex}) and the corresponding  long-time $\delta$-kick approximation (\ref{eq:tauTIapp}). Due to the effectively exponential expansion assisted by the inverted trap, the state at the time of the kick is highly accelerated with respect to the case under free expansion, and the discrepancy between the instantaneous and finite pulse approaches is more pronounced.

 Before closing this section, we note that while we have limited the discussion to protocols involving two stages between $\om_0$ and $\om_F$, time-optimal solutions can be found by increasing the number of intermediate jumps \cite{Stefanatos10}, paving new avenues to further extend the generalized DKC presented here.

\section{Discussion and conclusions}

We have reformulated DKC using the exact self-similar dynamics of quantum gases, which may be interacting and confined in a time-dependent harmonic trap, i.e., without resorting to the conventional classical description for noninteracting point-like particles. 
In doing so, we have found the exact condition for the strength of the  $\delta$-kick required to suppress (phase) excitations in an arbitrary scale-invariant evolution.
This leverages the use of  DKC beyond free flight, making it applicable in other self-similar expansions (and compressions)  in time-dependent traps. In addition, this allows one to consider control protocols based on a combination  of time-dependent traps and pulses.

We have further established the equivalence between the resulting generalization of  DKC and time-optimal bang-bang protocols derived from Pontryagin's maximum principle \cite{Salamon09,Stefanatos10} in the limit in which the frequency of the finite-time pulse is unbounded.
This equivalence has been shown with and without trap inversion.
This analysis also allows us to establish the conditions for the instantaneous pulse description in terms of $\delta$-kicks to hold, by estimating the leading corrections for finite-time pulses.

An application of the generalized framework for DKC in scale-invariant processes concerns the engineering of fast driving protocols from one equilibrium state to another.
Diabatic excitations under scale-invariance can occur due to the presence of  (i) a mismatch between the width of the expanding cloud and the target  equilibrium state and
(ii) a spatially varying phase across the extent of the atomic cloud.
Standard techniques to engineer STA in this context -including reverse-engineering scale-laws and counterdiabatic driving \cite{Chen10,Muga09,Stefanatos10,Hoffmann11,delcampo11,Choi2011,Choi2011b,delcampo12,Jarzynski13,delcampo13,Deffner14,Deng18,Deng18Sci,Diao18}- are limited by the need to suppress both kind of excitations at once. 
We have introduced a new class of STA assisted by kicks, in which the modulation of the trapping frequency suppresses type (i) while a final  $\delta$-kick  suppresses phase excitations of type (ii).
STA protocols assisted by kicks  ease the requirements on the magnitude of the required trapping frequency and allow for ultrafast expansions in which the width of the cloud grows exponentially as a function of time.
This class of STA can be readily generalized  beyond expansions and compressions of ultracold gases to other scale-invariant processes in which the application of kicks provides the means for control, e.g., in transport and rotation of atomic clouds and other matter waves, such as trapped ions. Likewise, further applications can be envisioned in process lacking scale invariance.

\begin{acknowledgments} 
It is a pleasure to acknowledge discussions with Aurelia Chenu and Fernando J. G\'omez-Ruiz. We acknowledge funding support from the Spanish MICINN (PID2019-109007GA-I00). DS \& AMS are supported by the Natural Sciences and Engineering Research Council (NSERC) of Canada; AMS is a Fellow of CIFAR.
\end{acknowledgments} 

\appendix

\section{Heisenberg uncertainty relation under scale invariance: DKC and STA}\label{AppendixHeis}

Consider  the Heisenberg uncertainty relation for an atomic cloud in a time-dependent harmonic trap.
From the definition of the scaling factor, denoting $R^2=\sum_{i=1}^{\N} \vec{r}_{i}\,^{2}$,  the variance in real space of an arbitrary state reads 
\beqa
\la R^2(t)\ra=b^2\la R^2(0)\ra, \quad \la R^2(0)\ra=\frac{\la H(0)\ra}{m\om_0^2}, 
\eeqa 
with the second identity being specific to scale-invariant systems at equilibrium.

To compute the variance in momentum space, we first recall that the time-evolution can be described by the action of the unitaries \cite{Lohe08,Jaramillo16,Beau20}
\beqa
\mathcal{S}&=&\exp\left[-i\frac{\log b}{2\hbar}\sum_{i=1}^\N (\vec{r}_i\cdot\vec{p}_i+\vec{p}_i\cdot\vec{r}_i)\right],\\
T_r&=&\exp\left[i\frac{m\dot{b}}{2\hbar b}\sum_{i=1}^\N \vec{r_i}^2\right].
\eeqa
Indeed, defining $T=T_r\mathcal{S}$,
\beqa
\Psi\left(\vec{r}_1,\dots,\vec{r}_\N,t\right)=e^{i\alpha_t}T\Psi_{n}\left(\vec{r}_1,\dots,\vec{r}_\N,0\right),
\eeqa
with $\alpha_t=-\int_{0}^t\frac{E(0)}{\hbar b(t')^2}dt'$ and $E(0)$ being the energy eigenvalue of $\Psi$ at $t=0$, i.e. $H(0)\Psi(0)=E(0)\Psi(0)$. However, the following treatment holds for mixed states as well.
It follows that $\mathcal{S}^\dag T_r^\dag\vec{p}_iT_r\mathcal{S}=\vec{p}_i/b+\vec{r}_i\dot{b}$ and thus
\beqa
 T^\dag\frac{1}{2m}P^2T=\frac{1}{2mb^2}P^2+\frac{\dot{b}}{b}C+\frac{m}{2}\dot{b}^2R^2
\eeqa
with $C =\frac{1}{2}\sum_{i=1}^{\N}(\vec{r}_i\cdot\vec{p}_i+\vec{p}_i\cdot\vec{r}_i)=\frac{1}{2}\sum_{i=1}^{\N}\left\{\vec{r}_i,\vec{p}_i\right\}$.
For an initial equilibrium state at $t=0$, $\la C(0)\ra=0$ and 
\beqa
\frac{1}{2m}\la P^2(t)\ra =\frac{1}{2mb^2}\la P^2(0)\ra+\frac{m}{2}\dot{b}^2\la R^2(0)\ra.
\eeqa

In a scale-invariant dynamics, the Heisenberg uncertainty relation at time $t$  reads
\beqa
\Delta R(t)\Delta P(t)=\Delta R(0)\sqrt{\Delta P(0)^2+m^2b^2\dot{b}^2\Delta R(0)^2},\nonumber\\
\eeqa
provided that $\la R(0)\ra=\la P(0)\ra=0$.
Only when $\dot{b}=0$, one finds 
\beqa
\Delta R(t)\Delta P(t)=\Delta R(0)\Delta P(0).
\eeqa
This is achieved by DKC, STA based on reverse engineering and adiabatic driving.
However, a generic nonadiabatic expansion  with $\dot{b}\neq 0$ leads to a suboptimal squeezing of the momentum distribution 
due to residual excitations associated with the phase modulations in the final state.

\section{Solution of the Ermakov equation \label{app:solerma}}

The Ermakov equation (\ref{eq:ermakov}) is a second order nonlinear differential equation which admits an exact solution  \cite{Pinney50}. 
To express it, consider the two independent solutions $u$ and $v$ of the linear equation $\ddot{b}+\om(t)^2b=0$,
satisfying $u(0)=b_0$, $\dot{u}(0)=\dot{b}_0$, $v(0)=0$, $\dot{v}(0)=1/b_0$. For a real constant frequency $\omega$, these fundamental solutions take the familiar form
\begin{eqnarray}
u(t)&=&b_0\cos(\omega t )+\frac{\dot{b}_0}{\omega}\sin(\omega t),\\
v(t)&=&\frac{1}{b_0\omega}\sin(\omega t).
\end{eqnarray}
Using them, the exact solution of the Ermakov equation reads \cite{Pinney50}, 
\begin{equation}
b(t)= \sqrt{\left(b_0\cos(\omega t)+\frac{\dot{b}_0}{\omega}\sin(\omega t)\right)^{2}+\left(\frac{\omega_{0}}{b_0\omega}\right)^{2}\sin^{2}(\omega t)}
\end{equation}
and satisfies $b(0)=b_0,\dot{b}(0)=\dot{b}_0$. 

If the boundary conditions are known at the end of the evolution $t=t_F$, one can  readily find the solution at earlier times. For example, assuming 
$b(t_{F})=b_{F}$ and  $\dot{b}(t_{F})=0$, and shifting the time variable  as $t\rightarrow t-t_F$, the scaling factor for  $t<t_F$ reads
\begin{eqnarray}
\label{eq:realfr}
b(t)=\sqrt{b^{2}_{F}+\left(\left(\frac{\omega_{0}}{b_{F}\omega}\right)^{2}-b^{2}_{F}\right)\sin^{2}(\omega(t-t_{F}))}.\nonumber\\
\end{eqnarray}

The above results hold for $\omega>0$.
In the case of the inverted trap with $\omega^{2}=-\omega^{2}_{I}<0$, the fundamental solutions of $\ddot{b}-\om_I^2b=0$ read
\begin{eqnarray}
u(x)&=&\frac{\dot{b}_0}{\omega_{I}}\sinh(\omega_{I} x)+b_0 \cosh(\omega_{I} x),\\
v(x)&=&\frac{1}{b_0\omega_{I}}\sinh(\omega_{I} x).
\end{eqnarray}
The square of the scaling factor is then given by
\begin{eqnarray}
b(t)^2&=&\left(\frac{\dot{b}_0}{\omega_{I}}\sinh(\omega_{I} t)+b_0\cosh(\omega_{I} t)\right)^{2}\nonumber\\
& & +\left(\frac{\omega_{0}}{b_0\omega_{I}}\right)^{2}\sinh^{2}(\omega_{I} t).
\end{eqnarray}
For the choice $b(0)=1,\dot{b}(0)=0$, one recovers the result \cite{Stefanatos10}
\begin{equation}
b(t)=\sqrt{1+\left(1+\left(\frac{\omega_{0}}{\omega_{I}}\right)^{2}\right)\sinh^{2}(\omega_{I} t)}.\nonumber\\
\end{equation}

\section{Pulse duration in optimal control \label{app:duration}}
The quantity
\begin{equation}
\alpha=\left(\frac{\dot{b}}{\omega_{0}}\right)^{2}+\frac{\omega^{2}}{\omega^{2}_{0}}b^{2}+\frac{1}{b^{2}}
\end{equation}
is a constant of motion for the solution of Ermakov equation (\ref{eq:ermakov}) since its derivative is proportional to Ermakov equation, i.e., it vanishes identically for any $t$ \cite{Stefanatos10}. It can thus be used to determine the duration of each of the intervals of the bang-bang protocol presented in Sec. \ref{sec:finite}.

Using the solutions of Ermakov equation denoted by $b_{1}$ in $(0,t_{k})$ and by $b_{2}$ in  $(t_{k},t_{k}+\tau_k)$, one can express this invariant as

\begin{eqnarray}
\label{eq:invcond}
\alpha=\left(\frac{\dot{b}_{1}}{\omega_{0}}\right)^{2}-\frac{\omega^{2}_{I}}{\omega^{2}_{0}}b^{2}_{1}+\frac{1}{b^{2}_{1}}&=&-\frac{\omega^{2}_{I}}{\omega^{2}_{0}}+1,\\
\alpha=\left(\frac{\dot{b}_{2}}{\omega_{0}}\right)^{2}+\frac{\omega^{2}_{k}}{\omega^{2}_{0}}b^{2}_{2}+\frac{1}{b^{2}_{2}}&=&\frac{\omega^{2}_{k}}{\omega^{2}_{0}}b^{2}_{F}+\frac{1}{b^{2}_{F}}.
\end{eqnarray}

Subtracting these two identities,  $b(t)$ can be determined at the time of the kick $t_{k}$,
\begin{eqnarray}
b(t_{k})=\sqrt{\frac{\omega^{2}_{k}b^{2}_{F}+\omega^{2}_{I}+\frac{\omega^{2}_{0}}{b^{2}_{F}}-\omega^{2}_{0}}{\omega^{2}_{k}+\omega^{2}_{I}}}.
\end{eqnarray}

Imposing the continuity of the scaling factor  $b_{1}(t_{k})=b(t_{k})$ and $b(t_k)=b_{2}(t_k)$ yields (\ref{eq:tkwI}) and (\ref{eq:taukwk2wI}). A similar derivation for $\omega_{I}=0$ yields (\ref{eq:t1t2}) and (\ref{etaukfp}).

\bibliography{QSLultracold_lib}

\end{document}